\documentstyle[aps]{revtex}
\begin{document}

\title{How short is too short? Constraining zero-range interactions in 
nucleon-nucleon scattering}

\author{Daniel R. Phillips and Thomas D. Cohen}

\address{Department of Physics, University of Maryland, College Park, MD, 
20742-4111,USA}

\date{\today}

\maketitle

\begin{abstract}
We discuss a number of constraints on the effects of zero-range
potentials in quantum mechanics.  We show that for such a potential $p
\cot(\delta)$, where $p$ is the momentum of the nucleon in the center
of mass frame and $\delta$ is the S-wave phase shift, must be a
monotonically decreasing function of energy. This implies that the
effective range of the potential is non-positive.  We also examine
scattering from the sum of two potentials, one of which is a
short-range interaction. We find that if the short-range interaction
is of zero-range then it must be attractive, and the logarithmic
derivative of the radial wave function at the origin must be a
monotonically decreasing function of energy. If the short-range
interaction is not of zero range then a constraint which gives the
minimum possible range for it to fit the phase shifts exists. The
implications of these results for effective field theory treatments of
nucleon-nucleon interactions are discussed.
\end{abstract}

\twocolumn

There has been much recent interest in the use of contact 
interactions in nuclear physics 
calculations~\cite{We90,We91,vK94,Or96,Co96A,Pa95,Be95,Ka96,Fr96}. 
This interest has been motivated by the possible application of effective field
theory (EFT) techniques to problems in nuclear physics.
In particular, it has been
argued that if one is not interested in details of the short-range 
piece of the nucleon-nucleon potential,
then a contact interaction and derivatives of contact
interactions may be used to parameterize it.
This parameterization may be thought of as the result of 
integrating out some or all of the
short-distance degrees of freedom of the theory
and then expanding the resulting nonlocal Lagrangian 
in terms of contact interactions, i.e., delta functions 
and derivatives of delta functions. For processes at
energies well below the scale of the nonlocality this approach has been 
thought to provide a good description of the physics of the nuclear system. 

In the scheme proposed by Weinberg~\cite{We90,We91} the contact
interactions are to be inserted as potentials into a nonrelativistic
Schr\"odinger equation. Since the Schr\"odinger equation with a
delta-function potential is not well defined, a regularization scheme
must be specified. Up until now two different types of regularization
have been used. van Kolck {\it et al.} chose a momentum-space cut-off
in their EFT studies of NN scattering~\cite{Or96}. On the other hand,
the recent work of Kaplan {\it et al.} used dimensional regularization
and the $\overline{{\rm MS}}$ renormalization scheme to calculate NN
scattering in the ${}^1S_0$ channel~\cite{Ka96}. In any perturbative
calculation the two methods of regularization would be equivalent---at
least to the prescribed order of the calculation.  However, in EFT
calculations of NN scattering the regularization scheme is used to
generate a potential which is then iterated to all orders via a
Schr\"odinger equation. Thus it is {\it not} clear that the two
regularization methods are equivalent in this problem.

In this paper we consider only the former type of regularization.  We
show that if one attempts to take the momentum-space cut-off to
infinity in such a scheme certain observables in NN S-wave scattering
{\it cannot} be reproduced.  Moreover, if a long-range interaction is
also present the physical effects of contact interactions regulated in
this way are still restricted.  These constraints demonstrate that
zero-range interactions cannot be used to parameterize the effects of
unresolved short-distance physics.

Here we define a zero-range interaction to be the 
limit of some set of finite-range interactions as some range
parameter, $R$, tends to zero.  Consider such a set of two-body,
time-reversal invariant (i.e., real and symmetric) potentials, 
$V_R(r,r')$. Note that $V_R$ must be nonlocal if in the limit 
$R \rightarrow 0$ it is to approach interactions including 
derivatives of delta functions.
For simplicity, in most of our derivations we will
assume $V_R(r,r') =0$ for all $r,r' \geq R$. In fact, this 
restriction  is more stringent than necessary and our results hold more
generally. The theory is now unambiguously defined by calculating
observables for arbitrary $R$ and then taking the limit $R \rightarrow 0$ 
at the end of the problem.  A zero-range interaction is called
``trivial'' if all its effects vanish when this limit is taken.
Nontrivial zero-range interactions require
the strength of the potential $V_R$ to diverge as $R$ goes to zero
in such a way that the physical effects of the potential remain finite.  

The possibility of a zero-range interaction not being able to
reproduce the effects of a general short-range potential was discussed
in Ref.~\cite{Co96B}.  There an extreme EFT where {\it all} exchanged
particles are integrated out was considered. The resulting Lagrangian
contains only contact interactions. Such a Lagrangian corresponds to
the very low-energy limit of a theory of nuclear interactions and was
discussed by Weinberg~\cite{We90,We91}.  In Ref.~\cite{Co96B} it was
shown that in such theories the only nontrivial zero-range
interactions are attractive.  Indeed, it has been known for some time
that repulsive zero-range interactions in nonrelativistic quantum
mechanics are necessarily trivial~\cite{Fr76,BF85}.  The physics
underlying this is quite simple. The most repulsive interaction with
compact support is a hard core. But, a hard core of radius $R$ merely
imposes the boundary condition that the radial wave function vanishes
at $R$.  However, the free radial wave function vanishes at $r=0$, so
as $R \rightarrow 0$ the hard core has no effect.  Since the effects
of any other repulsive interaction must be smaller than those of a
hard core it follows that repulsive zero-range interactions are
trivial.

This inability to describe repulsion when all exchanged particles are
integrated out does not necessarily mean that zero-range interactions
cannot be used in EFT approaches to the NN interaction.  Firstly,
low-energy nuclear interactions are attractive. Secondly, EFT
treatments of the NN force need not, and generally do not, integrate
out the pions. However, in this paper we will show that the result of
Ref.~\cite{Co96B} is not the only constraint on the effects zero-range
interactions can have on S-wave scattering observables.  Our
discussion is split into two parts; first we deal with theories in
which only a nontrivial zero-range interaction is present, and second
we discuss a theory with both a zero-range interaction and a
long-range potential. In both cases we derive constraints which
indicate that the use of zero-range interactions in descriptions of NN
scattering is highly problematic.

In the first case, where the only interactions are
of zero-range (including zero-range terms with an arbitrary number
of derivatives),
there is a very strong constraint on $p \cot (\delta)$, where $\delta$
is the S-wave phase shift. The function $p \cot (\delta)$ 
must be monotonically decreasing with energy:
\begin{equation}
\frac{d}{d p^2} \left [ p \cot (\delta ) \right ] \leq 0.
\label{eq:negdvtv}
\end{equation}
This in turn implies a negative effective range, $r_e$, since
\begin{equation}
\frac{r_e}{2} \,  \equiv \, \left. \frac{d}{d p^2}
\left [ p \cot (\delta ) \right ] 
\right|_{p^2 =0}.
\label{redef}
\end{equation}
This result is in direct conflict with the data for nucleon-nucleon
scattering since the $^1S_0$ phase shifts have a positive effective 
range~\cite{St93}.

This raises an interesting issue concerning regularization and
renormalization in the EFT approach.  In Ref.~\cite{Ka96} a theory
with a delta function interaction and two derivatives thereof is
considered.  Dimensional regularization and $\overline{\rm MS}$
renormalization are used to calculate two-body scattering, and an
expression for the $T$ matrix where the scattering length and
effective range are arbitrary input parameters related to bare
parameters of the EFT by renormalization conditions is obtained.
Nothing in this treatment restricts $r_e$ to be negative.  This
contradicts the general result (\ref{eq:negdvtv}) and raises questions
about the use of dimensional regularization and $\overline{\rm MS}$
renormalization.  At the very least it shows that the T-matrix
obtained in Ref.~\cite{Ka96} cannot be written as the limit of a
sequence of solutions to Lippmann-Schwinger equations with
finite-range interactions. It follows that the two methods which have been 
used so far to regulate the Schr\"odinger equation for NN scattering in EFT
are {\it not} equivalent.
 
Equation (\ref{eq:negdvtv}) follows straightforwardly from a very
general bound on
derivatives of phase shifts  obtained more than 40 years ago by 
Wigner~\cite{Wi55}. In particular,
Wigner demonstrated that for potentials which are identically zero
beyond some radius $R$:
\begin{equation}
\frac{d \delta}{d p} \, \geq \, - \, R \, + \, \frac{1}{2
p}\sin(2\delta + 2 p R).
\label{eq:Wignerbound}
\end{equation}
It is a simple exercise to rewrite this as
\begin{equation}
\frac{d}{dp^2} (p \cot(\delta)) \leq 
\left(\frac{R}{2\sin^2(\delta)} - \frac {\cos(2 \delta + p R) \sin (p R)}
{2 p \sin^2(\delta)} \right).
\label{eq:pcotdeltacons}
\end{equation}
Taking $R \rightarrow 0$ in Eq.~(\ref{eq:pcotdeltacons}) immediately yields
Eq.~(\ref{eq:negdvtv}), thus showing that
a theory with only zero-range interactions must have a
non-positive effective range.

Wigner derived the bound (\ref{eq:Wignerbound})
from general principles of causality and unitarity.
Below we present an alternative derivation of Wigner's bound, since the 
same technique can, with minor
modifications, be used to study the class of problems 
in which the potential includes both a long-range and a short-range 
interaction.

The radial Schr\"odinger equation for a central, nonlocal potential $V$ is:
\begin{equation}
-\frac{1}{2 \mu} \frac{d^2 u(r)}{d r^2}  + 4 \pi 
\int_0^\infty r V(r,r') r' u(r') dr'=Eu(r),
\label{eq:NLRSE}
\end{equation}
where the hermiticity of $V$ implies that $V(r,r')=V(r',r)$.
 If $V$ is time-reversal invariant, then the
radial wave function, $u(r)$, may be chosen to be real.

Define $u_i$, $i=1,2$, to be solutions of Eq.~(\ref{eq:NLRSE}) for two
different energies $E_1$ and $E_2$. Multiplying the equation for $u_1$ by
$u_2$ and vice versa, and integrating out to some radius ${\cal R}$
beyond which $V(r,r')$ is zero, yields
\begin{eqnarray}
\left. \left(u_2 \frac{du_1}{dr} - u_1 \frac{du_2}{dr}\right) 
\right|_{0}^{{\cal R}}=(p_2^2 - p_1^2) \int_0^{{\cal R}} dr u_1(r) u_2(r).
\label{eq:diff}
\end{eqnarray}
Next consider the wave functions
\begin{equation}
v_i(r)=\frac{\sin(p_i r + \delta(p_i))}{\sin(\delta(p_i))},
\label{eq:vdef}
\end{equation}
where $\delta(p)$ is the phase shift for scattering from $V$ at momentum 
$p=\sqrt{2 \mu E}$. Equation (\ref{eq:diff}) also applies to the $v_i$'s.
The difference of the 
equation for the $u_i$'s and that for the $v_i$'s may thus be taken 
and the result integrated out to radius ${\cal R}$.
Noting that $u$ may be normalized such that
$u_i({\cal R})=v_i({\cal R})$ then gives
\begin{eqnarray}
&& p_2 \cot(\delta(p_2)) - p_1 \cot(\delta(p_1))=\nonumber\\
&& \qquad (p_2^2 - p_1^2) \int_0^{{\cal R}} \, dr \, 
\left[v_1(r) v_2(r) - u_1(r) u_2(r) \right],
\label{eq:NLREeq}
\end{eqnarray}
where the boundary condition $u(0)=0$ has been used.

The derivation of Eq.~(\ref{eq:NLREeq}) for the special case of
local potentials 
may be found in standard scattering theory texts~\cite{GW64,Ne82}.
Choosing $p_1=0$ and $p_2=p$ then leads to:
\begin{equation}
p \cot (\delta)=-\frac{1}{a} + p^2 \int_0^{{\cal R}} \, dr \, 
\left[v_0(r) v_E(r) - u_0(r) u_E(r) \right],
\label{eq:reexp}
\end{equation}
where $a$ is the scattering length, assumed here to be nonzero.
Equation (\ref{eq:reexp}) is the basis for the effective range expansion. 
Now, if $V(r,r')=0$ for $r,r' \geq R$ with $R$ finite then 
\begin{equation}
r_e=2 \int_0^R \, dr \,[v_0^2(r) - u_0^2(r)].
\end{equation}
But, since
\begin{equation}
v_0(r)=\lim_{p \rightarrow 0} \frac{\sin(p r + \delta)}{\sin(\delta)}
= 1 - \frac{r}{a},
\end{equation}
and $u$ is real so $u^2 \geq 0$, it follows that
\begin{equation}
r_e \leq 2 \left[ R - \frac{R^2}{a} + \frac{R^3}{3 a^2} \right].
\label{eq:reconst}
\end{equation}
Thus, as the limit of a zero-range interaction is approached the
effective range must be non-positive---no matter what sequence of
potentials is used in approaching this limit. Note also that if the
effective range is positive then there is some minimum range the
potential can have if it is to produce phase shifts which agree with
experiment.

Furthermore, by rearranging (\ref{eq:NLREeq}), taking the limit as $p_2$
goes to $p_1$, and then using Eq.~(\ref{eq:vdef}) for $v$
and the bound $u^2 \geq 0$, we obtain Eq.~(\ref{eq:pcotdeltacons}),
which is completely equivalent to Wigner's bound (\ref{eq:Wignerbound}).

It is worth noting that the condition that the potential be identically
zero beyond some radius $R$ is more restrictive than necessary.  
For example, suppose $V$ has some characteristic
fall-off scale, $R$, which is to be taken to zero to obtain the limiting case
of a zero-range interaction. Provided that $u$
approaches the asymptotic solution $v$ sufficiently quickly, our constraint
on $p \cot (\delta )$
will still hold as $R \rightarrow 0$. In particular, if for each
$p$ there exists a positive real number $\epsilon$ and a real number $C$ 
(which may depend on $\epsilon$), such that, for all $r > R$,
\begin{equation}
u_p^2(r)=v_p^2(r) + C \epsilon (\frac{R}{r})^{1 + \epsilon}, 
\label{eq:condition}
\end{equation}
then upon taking the limit as ${\cal R} \rightarrow \infty$ we find
\begin{equation}
\int_0^{\infty} \, dr \, [v_p^2(r) - u_p^2(r)]
\leq R(1 - C) + {\cal O}(R^2).
\end{equation}
Thus, as $R \rightarrow 0$ we recover the result (\ref{eq:negdvtv}).

These results suggest that nuclear phenomena cannot be sensibly described
using zero-range interactions alone. However, this is hardly a new observation.
In the 1930's it was argued on the basis of a variation-al principle
that two-body zero-range interactions yielding a bound deuteron imply an
infinitely bound triton~\cite{Th35,Be36}. This effect was explicitly 
seen in three-body calculations in the 1970's~\cite{Gi73}.  

Such difficulties in theories with zero-range
interactions alone are problems of principle for EFT calculations in 
nuclear physics. However, they
need not arise in practice, since pions are usually explicitly included.
Therefore, a total interaction which is the sum of a regulated contact
interaction, $V_R(r,r')$, and a 
long-range potential, $V_L(r)$, is of more interest.
For concreteness we consider the case
where $V_L(r)$ is the longest-range piece of one-pion exchange, i.e.,
\begin{equation}
V_L(r)=-\alpha_\pi \frac{e^{-m_\pi r}}{r}.
\end{equation}
The results derived here go through---with minor modifications---for any 
potential which  is sufficiently non-singular as $r \rightarrow 0$.
The nonlocal radial Schr\"odinger equation is:
\begin{equation}
-\frac{1}{2 \mu} \frac{d^2u}{dr^2} + V_L(r) u(r) + 
\int r V_R(r,r') r' u(r') \, dr'=E u(r).
\label{eq:SEwithVL}
\end{equation}
The following discussion of this equation is 
based on the one in Appendix A of Ref.~\cite{Ka96}.

For $r > R$ there will be two solutions to Eq.~(\ref{eq:SEwithVL});
one which is
regular in the limit $r \rightarrow 0$, which we denote $J_E(r)$, and one
which is irregular in that limit, which we denote $K_E(r)$. Their asymptotic
behavior is: as $r \rightarrow 0$
\begin{eqnarray}
J_E(r) &\rightarrow& r - \alpha_\pi \mu r^2 + {\cal O}(r^3), 
\label{eq:Jsmallr}\\
K_E(r) &\rightarrow& \frac{\mu}{2 \pi} - \frac{\alpha_\pi \mu^2}{\pi} r
\ln (\lambda r) + {\cal O}(r^2 \ln(r)), \label{eq:Ksmallr}
\end{eqnarray}
(where $\lambda$ is arbitrary); and, as $r \rightarrow
\infty$
\begin{eqnarray}
J_E(r) \rightarrow \tilde{y} e^{ipr} + \tilde{y}^* e^{-ipr},
K_E(r) \rightarrow \tilde{z} e^{ipr} + \tilde{z}^* e^{-ipr}.
\end{eqnarray}
In the free case ($\alpha_\pi=0$) $\tilde{y}=\frac{1}{2ip}$. Generally, the 
values of $\tilde{y}$ and $\tilde{z}$ are related by the constraint:
\begin{equation}
\tilde{y} \tilde{z}^* - \tilde{y}^* \tilde{z}=- \frac{i \mu}{4 \pi p}.
\end{equation}
For $r > R$ the solution of Eq.~(\ref{eq:SEwithVL}) is
\begin{equation}
u_E^+(r)=a J_E(r) + b K_E(r). 
\end{equation}
The coefficients $a$ and $b$ are to be determined by matching the logarithmic
derivative
\begin{equation}
\gamma_E(r)=\frac{d}{dr} \log(u_E(r))
\end{equation}
of the wave $u_E^+(r)$, which we denote $\gamma^+$, and 
that of the solution of Eq.~(\ref{eq:SEwithVL}) for $r < R$, $u^-(r)$, which 
we denote $\gamma^-$, at the ``matching point'' $r=R$.
Direct evaluation from the expansions (\ref{eq:Jsmallr}) and (\ref{eq:Ksmallr})
gives, for small $R$,
\begin{equation}
\gamma^+(R) = \frac{2 \pi}{\mu} \frac{a}{b} - 2 \alpha_\pi \mu
\ln(\lambda R) - 2 \alpha_\pi \mu + {\cal O}(R).
\label{eq:gamma+R}
\end{equation}
Note that unless $\alpha_\pi=0$, $\gamma^+(R)$ diverges as $R \rightarrow 0$. 

As a specific example, suppose the potential $V_R(r,r')$ is a surface 
delta function, i.e., choose
\begin{equation}
V_R(r,r')=\frac{C}{4 \pi R^2} \delta(r - R) \delta^{(3)}(r - r').
\label{eq:SDI}
\end{equation}
By integrating 
Eq.~(\ref{eq:SEwithVL}) from $R - \eta$ to $R + \eta$, where $\eta$ is 
a small positive infinitesimal, we find:
\begin{equation}
4 \pi R^2(\gamma^+(R) - \gamma^-(R))=C.
\label{eq:contcond}
\end{equation}
For $r < R$ the solution of the radial Schr\"odinger equation
(\ref{eq:SEwithVL}) with the potential (\ref{eq:SDI}) is
$u_E^-(r)=a' J_E(r)$.
Calculating the logarithmic derivative $\gamma^-(R)$ from the small-$r$ 
expansion (\ref{eq:Jsmallr}) and then substituting the result and 
Eq.~(\ref{eq:gamma+R}) into Eq.~(\ref{eq:contcond}) then yields, for small $R$,
\begin{equation}
\frac{2 \pi}{\mu} \frac{a}{b} - 2 \alpha_\pi \mu
\ln(\lambda R) - \alpha_\pi \mu + {\cal O}(R)=\frac{1}{R} + 
\frac{C}{4 \pi R^2}.
\end{equation}
Since $a/b$ is a finite ratio, taking the limit as $R \rightarrow 0$ 
shows $C \rightarrow - 4 \pi R$
as $R \rightarrow 0$, and so we see that the short range potential must
be attractive in that limit. 

This generalizes the result discussed in Ref.~\cite{Co96B}:
If physical observables are to be kept finite
it is impossible to construct a sequence of potentials whose
limit is a repulsive zero-range interaction. Observe that ultimately the 
details of the
long-range potential $V_L(r)$ were not used in our obtaining this result. 
As in the $\alpha_\pi=0$ case of Ref.~\cite{Co96B}, the result should also be 
independent of the particular set of potentials chosen for
$V_R(r,r')$.

We now turn to the derivation of an equation analogous to 
Eq.~(\ref{eq:negdvtv}). Our argument is completely
independent of the choice of the short range potential $V_R(r,r')$.
Although the logarithmic derivative, $\gamma^+(R)$, is generally
not well behaved as $R \rightarrow 0$, by
evaluating $\gamma^+(R)$ at two energies $E_2$ and $E_1$ and
taking the difference we eliminate these divergent terms. In other words,
\begin{equation}
\gamma_{E_2}^+(R) - \gamma_{E_1}^+(R)
=\frac{2 \pi}{\mu} \left[\frac{a(E_2)}{b(E_2)} - \frac{a(E_1)}{b(E_1)}\right] 
+ {\cal O}(R).
\label{eq:gammaabreln}
\end{equation}
Now choose $v_E$ to be the analog of the free solution, 
i.e., the solution of the Schr\"odinger equation (\ref{eq:SEwithVL}) with 
$V_R$ set equal to zero: $v_E(r)=u_E^+(r)/u_E^+(0)$.
Then following exactly the arguments given above for the case 
$\alpha_\pi=0$ gives,
\begin{eqnarray}
\frac{d}{d E} \left(\left.\frac{d v_E(r)}{dr}\right|_{r=R} \right)
\leq M \int_{0}^R dr \, v_E^2(r).
\label{eq:EffrangeVL}
\end{eqnarray}
This is a constraint which must hold regardless of the value of $R$.
But, since in the limit $R \rightarrow 0$ 
$\left. \frac{d v_E(r)}{dr} \right|_{r=R}$ tends towards $\gamma_E(R)$,
Eq.~(\ref{eq:EffrangeVL}) implies that
\begin{equation}
\lim_{R \rightarrow 0} \frac{d\gamma_E(R)}{dE} \leq 0.
\label{eq:neggamma}
\end{equation}
As $\gamma_E$
is the logarithmic derivative of the radial Schr\"odinger wave function,
in the case $\alpha_\pi=0$ this reduces to Eq.~(\ref{eq:negdvtv}).

Moreover, in the limit $R \rightarrow 0$ $\frac{d\gamma_E(R)}{dE}$
may be related to $\frac{d}{dE} \left(\frac{a(E)}{b(E)}\right)$
using Eq.~(\ref{eq:gammaabreln}).
The total phase shift is then given by~\cite{Ka96}:
\begin{equation}
e^{2 i \delta}=e^{2 i \delta_\pi} - \left(\frac{1}{\tilde{y}^*}\right)^2
\frac{i \mu}{4 \pi p} \frac{1}{a/b + \tilde{z}^*/\tilde{y}^*},
\end{equation}
where $\delta_\pi$ is the phase shift arising from $V_L$ alone, 
and $\tilde{z}$ and $\tilde{y}$ are determined by the asymptotic 
behavior of the solutions to Eq.~(\ref{eq:SEwithVL}). Thus in the limit
$R \rightarrow 0$ the constraint 
(\ref{eq:neggamma}) places strong restrictions on the behavior of 
$\delta(E)$, but these also involve the energy-dependence of
the functions $\delta_\pi$, $\tilde{y}$ and $\tilde{z}$.

Observe that the use of one-pion exchange as the long-range
potential, $V_L(r)$, was not crucial to the proof of Eq.~(\ref{eq:neggamma}).
The only conditions which the potential $V_L(r)$ must satisfy in order for 
Eq.~(\ref{eq:EffrangeVL}) to hold are:
\begin{enumerate}
\item The Schr\"odinger equation with $V_L(r)$ should have
two solutions: one which is regular as $r \rightarrow 0$ and one which
is irregular as $r \rightarrow 0$. 

\item The irregular solution should have a divergence which is no worse than
$\frac{1}{r}$.

\item All pieces of the irregular solution which are divergent in
the limit $r \rightarrow 0$ should have energy-independent coefficients. 
\end{enumerate}
The last two conditions will always be satisfied for any $V_L(r)$ for which
the Schr\"odinger equation can be solved. 

So, we see that even a ``realistic'' EFT calculation, in which pions
are explicitly included while all other particles are integrated out
and their effects expanded in terms of interactions which are strictly
of zero range must obey certain constraints.  Firstly, the zero-range
interactions must be attractive if they are to have any effect on the
physical observables at all.  Secondly, the logarithmic derivative of
the radial wave function at the origin must be a monotonically
decreasing function of energy.  While it is not as simple to relate
this logarithmic derivative to scattering observables as it is in the
case $V_L=0$, the constraint still places severe restrictions on how
the phase shifts vary with energy.  Furthermore, the result
(\ref{eq:EffrangeVL}) may be used to derive a minimum radius $R$ for
which the potential $V_R(r,r')$ produces phase shifts which fit the
${}^1S_0$ scattering data. Results from such a calculation will be presented 
in a forthcoming publication~\cite{Ph96}.

\acknowledgements{We thank Bira Van Kolck for useful discussions. 
We also thank Iraj Afnan and Ben Gibson for calling our attention to the
work of L.~W. Thomas.
This work was supported in part by the U.S. Department of Energy
under grant no. DE-FG02-93ER-40762.}

\end{document}